\documentclass[%
 aip,
 jmp,%
 amsmath,amssymb,
 reprint,%
unsortedaddress
]{revtex4-1}

\usepackage{graphicx}
\usepackage{dcolumn}
\usepackage{bm}
\usepackage{float}
\usepackage{gensymb}
\usepackage{color}

    \usepackage{amsmath}
    \usepackage{amssymb}
    \usepackage{amsthm}
    \usepackage{amsfonts}

\begin{document}
	\title{Tailoring optical and magnetic properties of  molybdenum disulphide nanosheets by incorporating plasmonic gold nanoparticles}
	
		\author{Anjali Rani} 
		\affiliation{Department of Physics, School of Basic and Applied Sciences, Maharaja Agrasen University, Baddi,  India}
		
		\author{Arun Singh Patel} 
		\affiliation{School of Computational \& Integrative Sciences, Jawaharlal Nehru University, New Delhi-110067, India}
	\affiliation{Department of Physics, Indian Institute of Technology Delhi, New Delhi-110016, India}
	
	\author{Anirban Chakraborti} 	
	\affiliation{School of Computational \& Integrative Sciences, Jawaharlal Nehru University, New Delhi-110067, India}
	\email{anirban@jnu.ac.in}
	
		\author{Kulvinder Singh} 	
		\affiliation{Department of Chemistry, School of Basic and Applied Sciences, Maharaja Agrasen University, Baddi, India}
		
				\author{Prianka Sharma} 
				\affiliation{Department of Physics, School of Basic and Applied Sciences, Maharaja Agrasen University,  Baddi, India}
				\email{royprianka04@gmail.com}


	\begin{abstract}
	Abstract: The present paper deals with the systematic growth of plasmonic gold nanoparticles (Au NPs) on molybdenum disulphide (MoS$_2$) nanosheets as well as the  effect of Au NPs on the optical and magnetic properties. The crystalline nature of the nanocomposites is confirmed by X-ray diffraction and transmission electron microscopic techniques. The optical properties are characterized using absorption and Raman spectroscopic techniques. The electron paramagnetic resonance technique is used to study the magnetic response of the nanocomposites. The paper attempts to gain a fundamental understanding of the two-dimensional nanomaterial-based composites for their applications in the magnetic and optical devices.        	
		
	\end{abstract}
	
	
	\maketitle
\section{Introduction}	
In recent years, transition metal dichalcogenides (TMDs) have gained significant interest   due to their vast applications in the field of energy transfer, catalysis, optical and electronic devices.\cite{ goodfellow2016, prins2014, latorre2015innovative,   wang2012, wu2018, patelrsc2017, rani2019visible} Among these TMD materials, molybdenum disulphide (MoS$_2$) and tungsten disulphide (WS$_2$) have been explored as various semiconducting devices. The bulk TMDs generally comprise many layers of  nanosheets. The weak van der Waals interaction in the inter-layer of TMDs binds them to form  bulk materials. The optical and electrical properties of TMDs  strongly  depend on the number of layers. The TMDs have a band gap that is dependent on the number of layers. In  bulk form, they have indirect band gap while in  case of monolayer,  a direct band gap is observed.\cite{mak2010atomically} The transition from indirect to direct band gap results in high photoluminescence from  TMDs nanosheets.\cite{chhowalla2013} Having a band gap in the visible region renders more usefulness of the TMDs nanosheets in opto-electronic devices as compared to graphene, where a zero band gap is observed.\cite{meric2008rf} In the monolayer of  TMDs there are three layers of atoms -- a layer of transition metal (Mo$/$W) sandwiched between two layers of chalcogenides (S$/$Se).\cite{chakrabortipla2016} Thus, it forms a MX$_2$ structure.  In  bulk form, MoS$_2$ has 1.2 eV band gap while the monolayer has 1.9 eV band gap. Similarly, WS$_2$ has 1.3 eV band gap in bulk and 2.1 eV in the single layer form.\cite{lee2012, chhowalla2013, arunapl2016} Owing to this characteristic, finite band gap MoS$_2$ monolayer can complement the weakness of gapless graphene,  making it a promising two dimensional (2D) material for next generation applications in switching and opto-electronic devices. Thus, MoS$_2$ finds its applications in diverse fields like sensors, energy storage in lithium-ion batteries, opto-electronic devices, flexible electronic devices, photoluminescence, valleytronics and field-effect transistors.\cite{zhu2013single, chen2017controllable, song2016high, zibouche2014transition, bang2014}  MoS$_2$   has also been explored as catalyst for hydrodesulphurization and hydrogen evolution in recent years.\cite{lukowski2013, luo2018, paredes2016, lee2013} Atomically-thin MoS$_2$ sheets have relatively large in-plane mobility  and mechanical stability. 

The physical and chemical properties of MoS$_2$ nanosheets can be manifested many fold by incorporating noble metal nanoparticles. Hence,  nanocomposites can be used in a vast area of applications.  Among  noble metal nanoparticles, gold (Au) nanoparticles have an advantage due to its non-toxic nature and excellent stability. Further, Au is known to have a strong affinity for sulphur, which has been exploited to enhance charge transportation along interplanar directions and act as spacers to inhibit re-stacking of the TMDs. Recently, various research groups have synthesized and used  Au-MoS$_2$ nanocomposites in photocatalysis, surface enhanced Raman scattering and photovoltaic applications. Few research groups have also worked in the field of synthesis of Au NPs decorated on MoS$_2$ sheets and its applications as hydrogen catalysis, Fermi level engineering, photoluminescence quenching, etc.\cite{song2016high, chen2017controllable, zibouche2014transition}

There are various studies on the optical and electrical properties of  metal based MoS$_2$ nanosheets.\cite{miao2015surface, su2015role, li2017superior, kaushik2014schottky}   Besides, there are a few reports on the magnetic properties of metal based two dimensional nanosheets, especially the electron paramagnetic resonance (EPR) studies. In the present paper,  we report  chemical exfoliation of MoS$_2$ nanosheets for spontaneous decoration of Au NPs on MoS$_2$ sheets. A direct redox reaction is done with a gold precursor chloroauric acid (HAuCl$_4$) and sodium dodecyl benzene sulfonate as the stabilizing agent. The Au NPs are successfully deposited on  chemically exfoliated sheets. The optical and magnetic properties of the nanocomposites are then explored using optical spectroscopic techniques and EPR spectroscopy.

\section{Experimental methods}
\subsection{Materials}
 Molybdenum disulphide (MoS$_2$) powder, N-Methyl 2-pyrrolidone (NMP), Isopropyl alcohol (IPA), Hydrogen tetrachloroaurate trihydrate (HAuCl$_4$.3H$_2$O), Sodium dodecyl benzene sulfonate (SDBS) were obtained from Sigma Aldrich. All the chemicals were used without further purification.  

\subsection{Chemical exfoliation of MoS$_2$ nanosheets}
 MoS$_2$ nanosheets have been synthesized via liquid phase exfoliation method. For the synthesis  of MoS$_2$ nanosheets, 0.05g of bulk MoS$_2$ powder was dispersed in 20 mL of NMP, IPA and distilled water (volume ratio 3:1:1) mixed solvent. The solution was ultrasonicated continuously for 24 hours at room temperature. After 24 hours of sonication the colour of solution changed from dark greyish black to greenish. The resultant green dispersion obtained was retained for 30 min for settling down the un-exfoliated MoS$_2$. The clear supernatant was collected and 10 mg of SDBS was added to it. The resultant solution was then allowed to settle down in the next 30 min. 
    
\subsection {Synthesis of Au-MoS$_2$ nanocomposites}
 For the synthesis of Au-MoS$_2$ nanocomposites, stock solution of 10 mM Au salt (HAuCl$_4$.3H$_2$O) was prepared. Four sets each of 2 mL dispersion of exfoliated MoS$_2$ nanosheets were taken for the further synthesis of Au-MoS$_2$ nanocomposites. In these four sets, different quantities of Au salt solution (2$\mu$L, 4$\mu$L, 6$\mu$L and 8$\mu$L) were added and the mixtures were kept for aging for 48 hours. 

\subsection{Characterization techniques} The physical properties of MoS$_2$ and Au-MoS$_2$ nanocomposites were characterized by using spectroscopic and microscopic techniques. The morphology of the nanomaterial was investigated using transmission electron microscopy (TEM). The TEM analysis of MoS$_2$ and Au-MoS$_2$ nanocomposites was done   using JEOL 2100F transmission electron microscope operated at 200 kV.  The crystallinity of MoS$_2$ and Au-MoS$_2$ nanocomposites was explored using X-ray diffractometer from PANalytical with Cu K$\alpha$ (0.15 nm) as X-ray source. The Raman spectra of MoS$_2$ and Au-MoS$_2$ nanocomposites were recorded by a confocal Raman microscope with an excitation source as 532 nm laser. The absorption spectra of MoS$_2$ and Au-MoS$_2$ were recorded using UV-Vis absorption spectrophotometer (Labtronics, LT-2700, India). The EPR spectra were obtained at room temperature using EPR spectrometer (Bruker, EMX MicroX). 
\section{Results and discussion}
\subsection{TEM}
\begin{figure}
	\centering
	\includegraphics[width=0.95\linewidth]{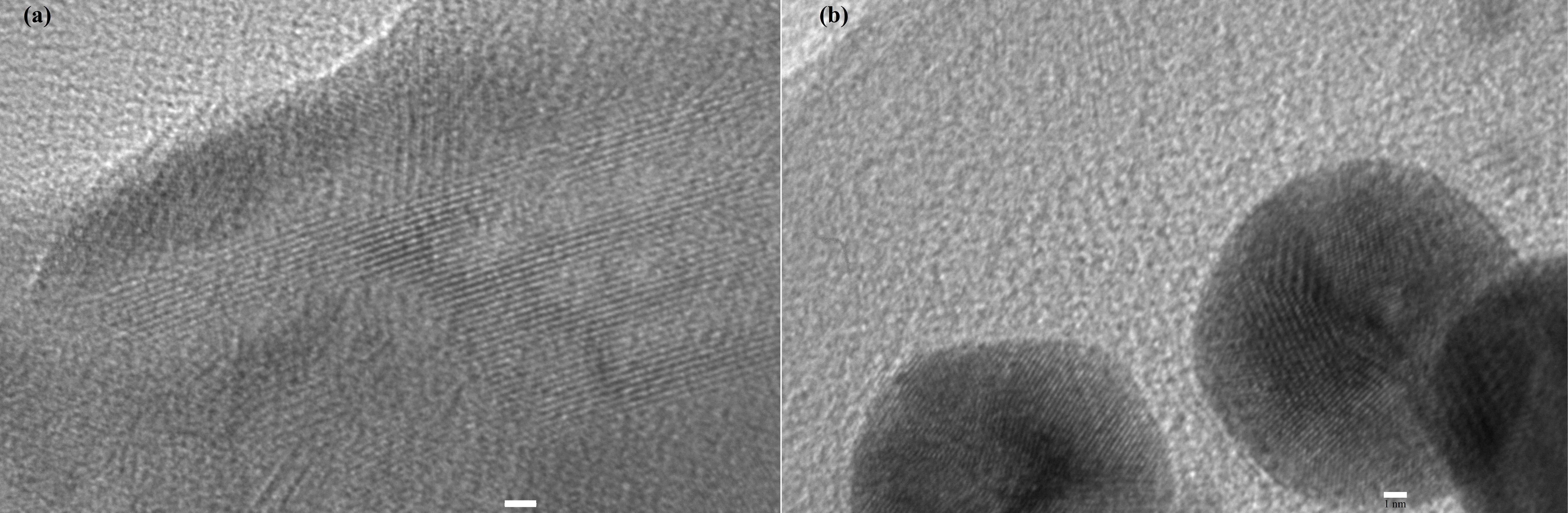} 
	\caption{TEM images of (a) MoS$_2$ and (b) Au-MoS$_2$ nanosheets; the white scale bar is equivalent to 1 nm. \label{TEM}}
\end{figure}

The shape and size of MoS$_2$ and that of Au NPs on the MoS$_2$ nanosheets were investigated using transmission electron microscopy (TEM) set-up. The TEM images of MoS$_2$ and Au-MoS$_2$ nanosheets are shown in Fig. \ref{TEM}. 
In the Fig. \ref{TEM}(a) sheet-like structure is observed, which corresponds to exfoliated MoS$_2$. In MoS$_2$, the lattice spacing is of the order of 0.27 nm, which corresponds to (100) plane.\cite{luo2016mos2} In case of Au-MoS$_2$, the Au NPs are found to be of spherical  shape and  size of these particles is around 20 nm (Fig. \ref{TEM}(b)). The lattice plane spacing in the Au NPs corresponding to FCC lattice of Au is observed and  spacing is of the order of 0.22 nm, which corresponds to (111) plane of gold lattice.\cite{lin2008selective}         
\subsection{XRD} 
\begin{figure}
	\centering
	\includegraphics[width=0.95\linewidth]{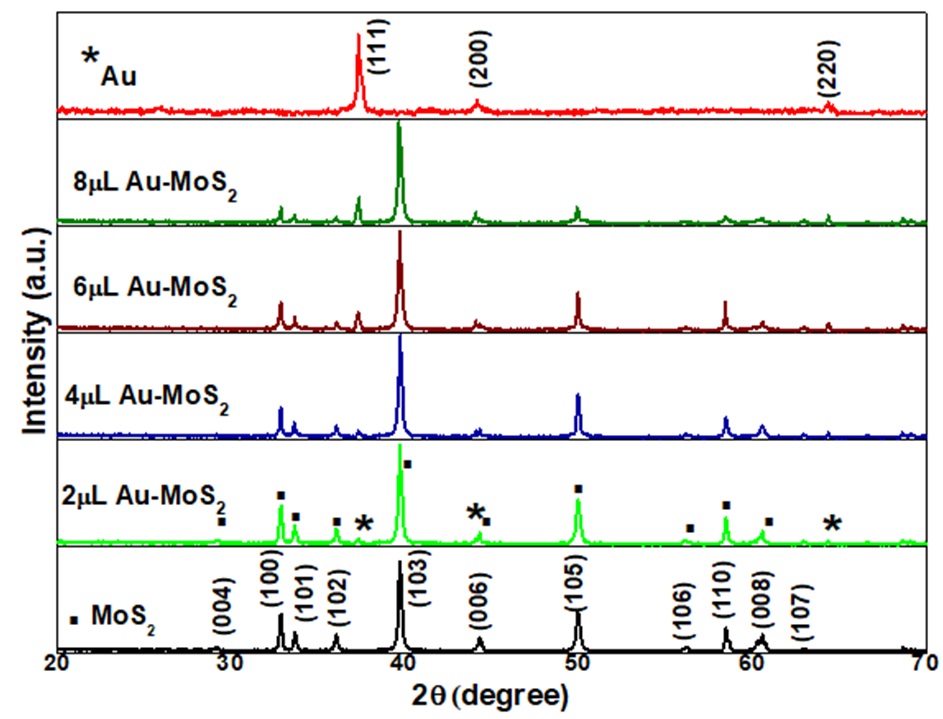} 
	\caption{XRD patterns  of  MoS$_2$ and Au-MoS$_2$  nanosheets with 2, 4, 6 and 8 $\mu$L   of Au-solutions along with pristine  Au and MoS$_2$. \label{XRD}}
\end{figure}

The XRD technique was used for investigating the crystal structure and lattice planes of the Au NPs and MoS$_2$ nanosheets. The XRD patterns of pristine Au and MoS$_2$  along with the Au-MoS$_2$ nanocomposites are shown in Fig.~ \ref{XRD}. In these XRD patterns, the diffraction peaks at 29.19$^{\circ}$, 32.88$^{\circ}$, 33.68$^{\circ}$, 36.09$^{\circ}$, 39.75$^{\circ}$, 44.34$^{\circ}$, 49.99$^{\circ}$, 56.19$^{\circ}$, 58.50$^{\circ}$, 60.61$^{\circ}$ and 62.95$^{\circ}$  are observed that are attributed to lattice planes (004), (100), (101), (102), (103), (006), (105), (106), (110), (008) and (107) of MoS$_2$, respectively    (JCPDS No. 37-1492).\cite{xu2015solvent} After Au NPs decoration, the Au-MoS$_2$ shows additional  diffraction peaks at 37.34$^{\circ}$, 44.17$^{\circ}$ and 64.40$^{\circ}$ corresponding to the (111) (200) and (220) planes, respectively, of Au phase (JCPDS No. 04-0784),\cite{jiang2012cu} that indicates  FCC Au NPs have been successfully decorated on the MoS$_2$ surface. Intensity of the  peaks corresponding to Au  enhances as  the concentration of Au increases in the nanocomposites, while the peaks of MoS$_2$ decrease correspondingly,   which is clearly observed in Fig. \ref{XRD}. Lowering  in  diffraction intensity is due to decrease in the exposure area of  MoS$_2$ nanosheets in presence of Au NPs. The presence of Au NPs hinders the exposure area of the MoS$_2$. 

\subsection{Raman study} 
\begin{figure}
	\centering
	\includegraphics[width=0.95\linewidth]{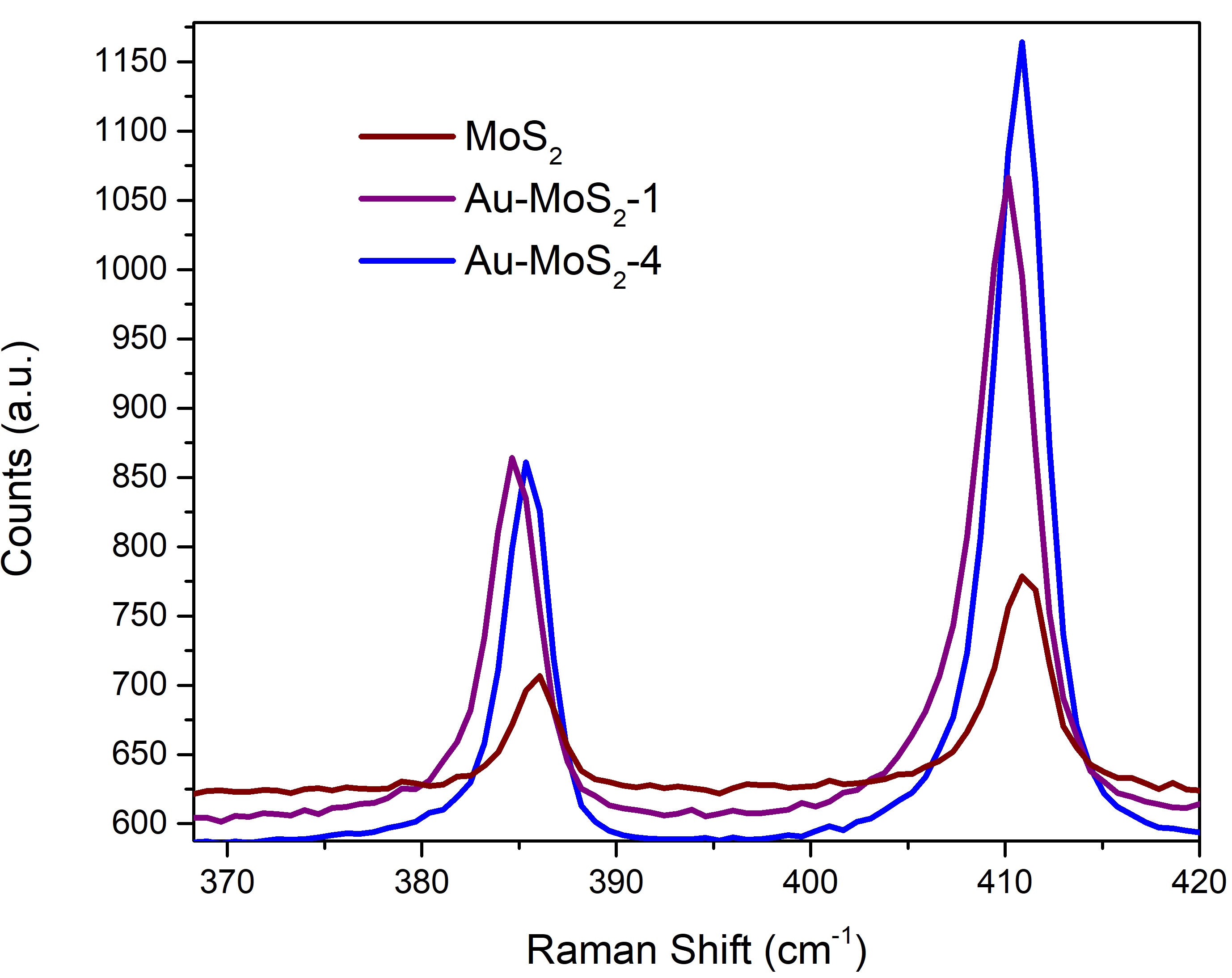} 
	\caption{Raman spectra of  MoS$_2$ and Au-MoS$_2$  nanosheets, where Au-MoS$_2$-1 is for 2 $\mu$L of Au-solution and Au-MoS$_2$-4 is for 8 $\mu$L of Au-solution. \label{Raman}}
\end{figure}

The vibrational modes of MoS$_2$ nanosheets and the effect of Au NPs on them were investigated using Raman spectrometer. The Raman spectra of MoS$_2$ and Au-MoS$_2$ nanosheets are shown in Fig. \ref{Raman}. The Raman spectra exhibit   two distinct Raman peaks -- one around 385.9 cm$^{-1}$ and the other around 410.9 cm$^{-1}$. These peaks are associated with E$^1_{2g}$ and A$_{1g}$ modes of MoS$_2$, respectively.\cite{arunapl2016} The E$^1_{2g}$ mode is attributed to the in-plane vibration of Mo and S atoms and this mode is sensitive to the built-in strain of 2D MoS$_2$.  The A$_{1g}$  mode is related to the out-of-plane vibration of S atoms, which is a reflection of the interlayer van der Waals interaction. For Au-MoS$_2$,  red shift in the  E$^1_{2g}$ and A$_{1g}$ modes are observed. The shifting is attributed to the effect of lattice strain due to the curvature of MoS$_2$ shell. The frequency difference of the E$^1_{2g}$ and A$_{1g}$ peaks comes out to be about 25 cm$^{-1}$ that renders exfoliation of few-layers MoS$_2$ sheets.\cite{plechinger2012raman}   In the presence of Au NPs, the intensity of Raman peaks are also enhanced. The enhancement is due to the plasmonic effect of  Au NPs. The presence of Au NPs causes enhanced electric field near the surface of Au NPs and hence the intensity of Raman signal is enhanced. This contribution is due to the effect of localized surface plasmon resonance (LSPR) of Au nanoparticle cores, typically called surface enhanced Raman scattering (SERS).\cite{su2014creating}   


\subsection{UV-Vis absorption} 
\begin{figure}
	\centering
	\includegraphics[width=0.95\linewidth]{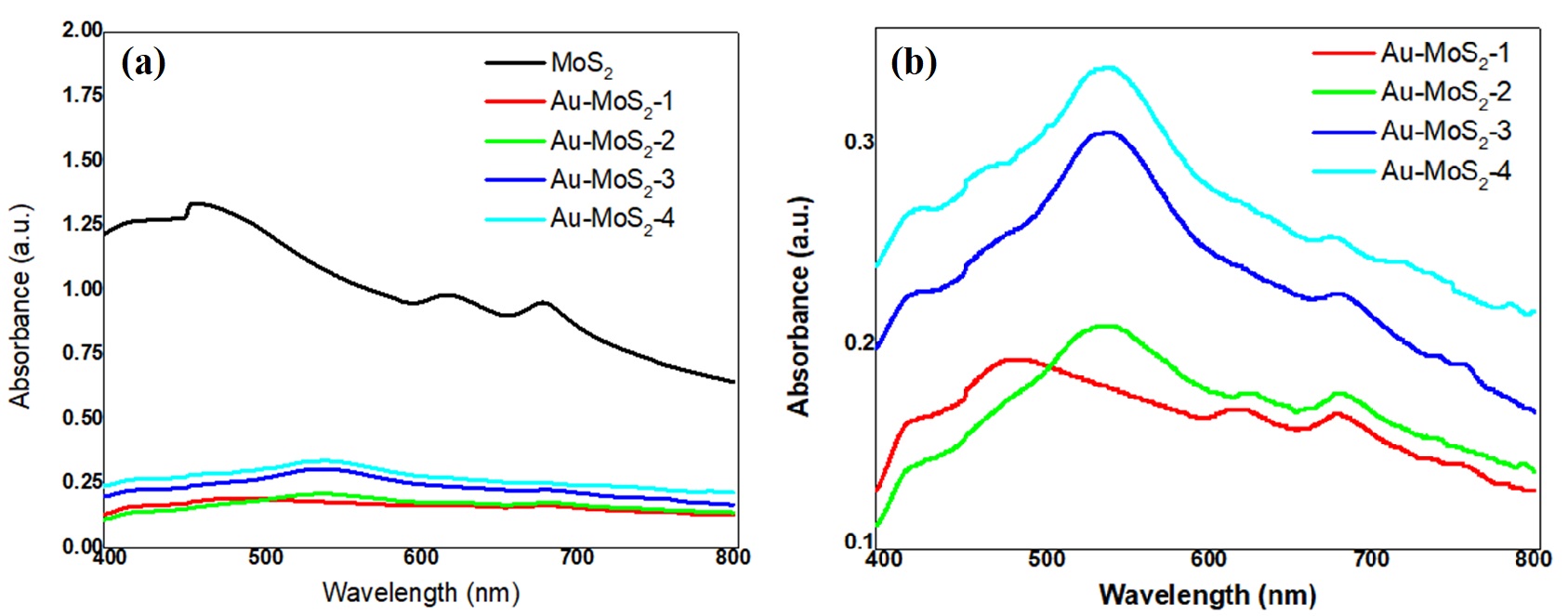} 
	\caption{(a) Absorption spectra of MoS$_2$ and Au-MoS$_2$  nanosheets with  2, 4, 6  and 8 $\mu$L    of Au-solutions shown as Au-MoS$_2$-1, Au-MoS$_2$-2, Au-MoS$_2$-3 and Au-MoS$_2$-4, respectively, and (b) shows the zoomed view of absorption arises due to presence of Au NPs in the Au-MoS$_2$ nanosheets \label{Abs}}
\end{figure}

The optical absorption of pristine MoS$_2$ nanosheets and Au-MoS$_2$ nanocomposites has been explored using UV-Vis absorption spectrophotometer. The absorption spectra of MoS$_2$ and Au-MoS$_2$ are shown in Fig.~ \ref{Abs}. The absorption spectrum of MoS$_2$ nanosheet shows two small humps in the visible range -- one at 620 nm and other at 670 nm, which are known as  B and A excitonic peaks, respectively.\cite{patelrsc2017} These peaks arise due to spin-orbit interaction causing splitting of valence band energy levels and the excitonic transitions between the splitted valence bands and the minima of conduction  band at the $K$-point of Brillouin zone.   The  interlayer coupling also plays an important role in the valence band splitting. When a small aliquot of gold precursor was added into chemically exfoliated MoS$_2$, a new absorption peak corresponding to the Au plasmon band at around 530 nm emerged, suggesting the consumption of Au$^{3+}$ and  formation of gold nanoparticles.\cite{mustafa2010surface} As we increase the Au concentration, the surface plasmon resonance peaks of Au experiences both red shift and increase in the intensity of peaks (shown in Fig.\ref{Abs}(b)). These shifting infer to strong plasmon-exciton coupling between Au and MoS$_2$. The surface plasmon resonance strongly depends on shape, size and separation of the nanoparticles along with the surrounding environment.\cite{amendola2017surface} In Fig.~\ref{Abs}(b), it is observed that when the concentration of Au is low the SPR peak is not so prominent, which is due to smaller size of Au NPs. When these particles have size less then 2-3 nm, the localized plasmons are not observed.   The extent of Au-ion reduction by the spontaneous redox reaction can be estimated by monitoring the absorption peak of Au$^{3+}$ and  quantifying the loading level of Au on the surface of MoS$_2$.


\subsection{Electron paramagnetic resonance study} 
\begin{figure}
	\centering
	\includegraphics[width=0.95\linewidth]{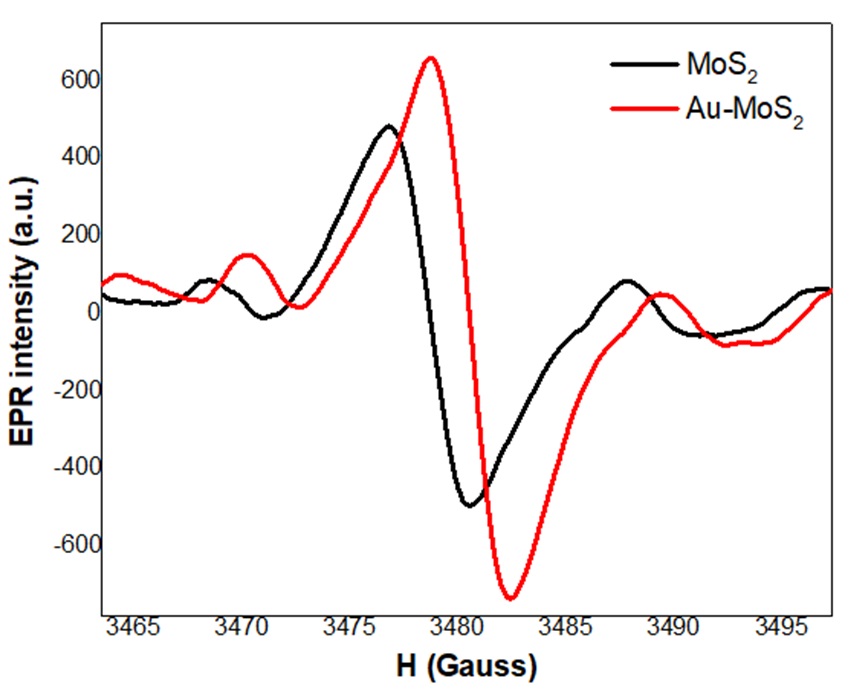} 
	\caption{EPR spectra of  MoS$_2$ and Au-MoS$_2$ nanosheets. \label{EPR}}
\end{figure}

The paramagnetic properties  of MoS$_2$ and Au-MoS$_2$ nanosheets are studied using electron paramagnetic resonance (EPR) spectroscopy. The EPR spectra of MoS$_2$ and Au-MoS$_2$ nanosheets are shown in Fig. \ref{EPR}. 
The EPR spectra originated due to the Mo−S dangling bonds, which are associated with different kinds of defects present in the MoS$_2$ structure.  The signature of the Mo−S dangling bonds can be detected as EPR absorbance around 3500 G of magnetic field and the intensity of the EPR peak  is  proportional to the concentration of Mo−S dangling bonds from the S-vacancies in the MoS$_2$ structure.\cite{cai2015vacancy} In case of pristine MoS$_2$,  an EPR signal at 3476 G  is observed, which shows that some defects are created during the chemical exfoliation process  using an ultrasonication bath. In case of Au-MoS$_2$ nanostructures, a significant increase in the intensity as well as slight shift in the EPR peak is observed (3478 G),  which is due to binding of Au with the sulphur atoms. The binding of Au-S leads to an overall increase in the sulphur vacant sites.\cite{cai2015vacancy}  Any kind of localized defect in the MoS$_2$ structure, such as  edge dislocation or  dangling bond, provides a paramagnetic Mo site, which is  characterized by an anisotropic g-tensor in the EPR.

\section{Conclusion}
In summary, the Au-MoS$_2$ nanocomposites have been synthesized by using chemical exfoliation and reduction method. Different characterizations revealed that these Au-MoS$_2$ nanocomposites show strong response towards the optical wavelengths due to the presence of surface plasmons in the nanocomposites, which enhances the charge transportation between MoS$_2$ nanosheets. The presence of plasmonic particles was found to alter the magnetic property of the nanocomposites. This study could be further extended to  other plasmonic nanoparticles of different shape and size, and their effect on the optical and magnetic response on the 2D nanomaterials could also be explored.

  Acknowledgment: AC is thankful to AIRF  JNU for TEM, Raman and EPR measurements, and AC is also thankful to DST-PURSE for partial financial support. Authors are thankful to Mr. Praveen Mishra for his help.

		\bibliography{main}

\end{document}